\documentclass[preprint,12pt]{elsarticle}

\usepackage{graphicx}
\usepackage{sidecap}
\usepackage{array}
\usepackage{subfigure}
\usepackage{color}
\usepackage{amsmath}
\usepackage{booktabs}
\usepackage{lineno,hyperref}

\newcommand{\bef}{\begin{figure}}
\newcommand{\eef}{\end{figure}}

\newcommand{\be}{\begin{equation}}
\newcommand{\ee}{\end{equation}}
\newcommand{\bea}{\begin{eqnarray}}
\newcommand{\eea}{\end{eqnarray}}

\journal{}

\begin{document}

\title{Constraining input parameters of AMPT model with $\phi$ meson production}


\author{Katyayni Tiwari and Md. Nasim}
\address{Department of Physics, Indian Institute of Science Education and Research, Berhampur-760010, India}




\begin{abstract}
 A Multi-Phase Transport Model (AMPT) has been extensively used to understand the physics behind the experimental observation.
Like other models, the outcome of the AMPT model depends on the initial parameters. Therefore,  one needs to choose suitable initial parameters before using the model. 
Lund string fragmentation function has been used to create quark-antiquarks pairs in the AMPT model. The Lund string fragmentation parameters determine the yields and transverse momentum ($p_{T}$) spectra of particles produced in nucleus-nucleus collisions. The values of  Lund string fragmentation parameters were determined by fitting the charged particle yield and  $p_{T}$ spectra measured in the experiment. 
In this paper, we have shown the yield of strange quarks carrying hadrons, e.g. $\phi$ mesons, are more sensitive to Lund string fragmentation parameters compared to non-strange pions.  Using $\phi$ meson spectra measured at RHIC, we have obtained new sets of Lund parameters for Au+Au collisions at $\sqrt{s_{\mathrm {NN}}}$ = 11.5, 39,  and 200 GeV.  We have found that using the same set of parameters, we can explain $\phi$-meson yield at $\sqrt{s_{\mathrm {NN}}}$ = 39  and 200 GeV, however, we need a different set of parameters for  $\sqrt{s_{\mathrm {NN}}}$ = 11.5 GeV. This suggests that at low energy, $\sqrt{s_{\mathrm {NN}}}$ = 11.5 GeV, the underlying mechanism for particle production is different compared to top RHIC energies. We have also predicted invariant yield of $\pi$ and $\phi$ mesons  as a function of $p_{T}$ in U+U collisions at $\sqrt{s_{\mathrm {NN}}}$ = 196 GeV to be measured by STAR experiment. 
 
 \end{abstract}


 \maketitle

 \section{Introduction}
 
The Relativistic Heavy Ion Collider (RHIC) has undertaken a  Beam Energy Scan program (BES) to look for a change in observations of various measurements as a function of
beam energy to establish the partonic phase at higher energy collisions~\cite{9gev,beswhite}. The first phase of the Beam Energy Scan program (BES-I) plus the top energy at RHIC has allowed access to a region of the QCD phase diagram covering a range of baryon chemical potential ($\mu_{B}$) from 20 to 420 MeV~\cite{beswhite}. 
In such a program, the energy dependence of $\phi (s\bar{s})$ meson production plays a crucial
role, since it has small hadronic interaction cross-section and freeze-out early compared to other non-strange hadrons~\cite{smallx,white,nuxu}.   
The first phase of the beam energy scan program has been completed and the results are being available to understand the properties of  produced matter created in heavy-ion collisions~\cite{chgv2_bes1,pidv2_bes1,pidprl_bes1,pidv1_bes1,chgsep_bes1,phi_omega_star,pikp_star,phenix1,phenix2,phenix3,phenix4}.
Various observables are compared to theoretical calculations to understand the physical mechanism behind the measurements~\cite{AuAuUU,v2model_nsm,v2phi_nsm,v2phi_nsm_bm,shusu1,shusu2,shusu3,bm1,bm2,bm3,bm4,cmko1,cmko2,cmko3,cmko4,fuq1,ziwi1,ziwi2,ziwi3}.   One of the frequently used models in heavy-ion collisions is AMPT~\cite{ampt}. The motivation of this paper is to find out suitable input parameters of AMPT model which explains the production of non-strange ($\pi$) and strange mesons ($\phi$) measured by the STAR experiment at top RHIC and BES energies and give a prediction for $\pi$ and $\phi$ meson $p_{T}$ spectra in U+U collisions at  $\sqrt{s_{\mathrm {NN}}}$ = 196 GeV.\\
The paper is organized as follows.  In the next section, we discuss the AMPT model. We also discuss the implementation of U+U collision in the AMPT model. Section 3 presents the results, which include the comparison of data and model calculation for $\pi$ and $\phi$ meson invariant yield as a function of $p_{T}$. Finally, in section 4 we present a summary of our findings.

\section{The AMPT Model}
 The AMPT uses the same initial conditions as in HIJING~\cite{hijing}. However, the minijet partons are made to undergo scattering before they are allowed
to fragment into hadrons.  In AMPT default model, hadrons are formed from these quarks and antiquarks by using a symmetric fragmentation function $ f(z) \propto z^{-1} (1-z)^{a} \exp(-b m_{T}^{2}/z)$, where $z$ denotes the light-cone momentum fraction of the produced hadrons with respect to that of the fragmentation string and $a$ and $b$ are free parameters~\cite{lund}. 
The string-melting (SM) version of the AMPT model (labeled here as AMPT-SM) is based on the idea that for energy densities beyond  a critical value of $\sim$ 1 GeV/$\rm {fm}^{3}$,
it is difficult to visualise the coexistence of strings (or hadrons) and partons. 
Hence the need to melt the strings to partons. This is done by converting the mesons to a quark and antiquark pair, baryons to three quarks, etc. 
The quark-antiquark pair production from  string fragmentation in the Lund model is based on the
Schwinger mechanism~\cite{schw_qqbar}. In the Schwinger mechanism, the production probability of the quark-antiquark pair is proportional to $\exp(-\pi m_{T}^{2}/k)$, where $m_{T}$ is the transverse mass of the quark and $k$ is the string tension, approximately as given by $k \propto 1/[b(2 + a)]$.  The AMPT model with string melting leads to hadron formation using a quark coalescence model.  The subsequent hadronic matter interaction is described by a hadronic cascade, which is based on a relativistic transport (ART) model~\cite{art}.\\
\bef
\begin{center}
\includegraphics[scale=0.38]{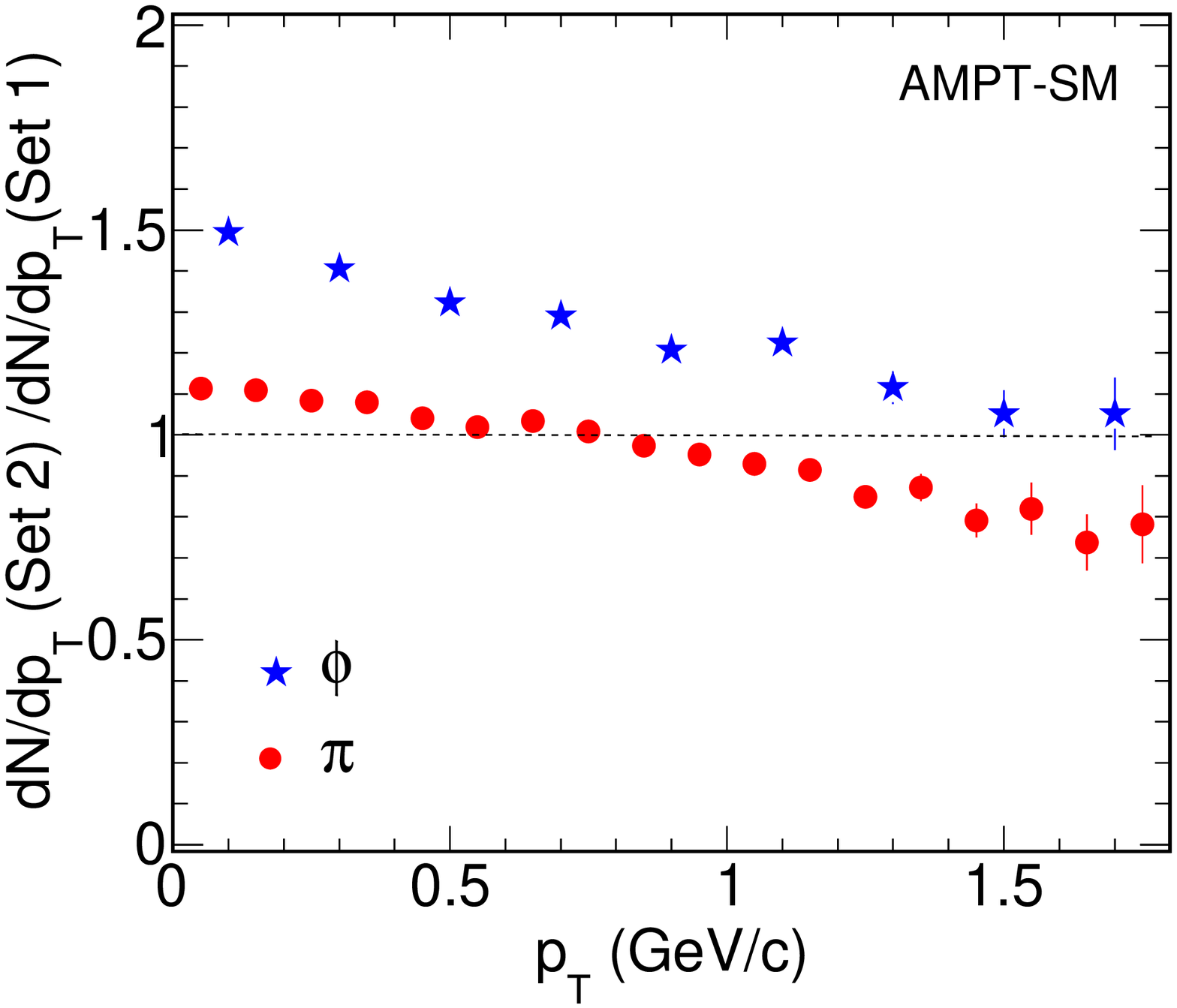}
\includegraphics[scale=0.38]{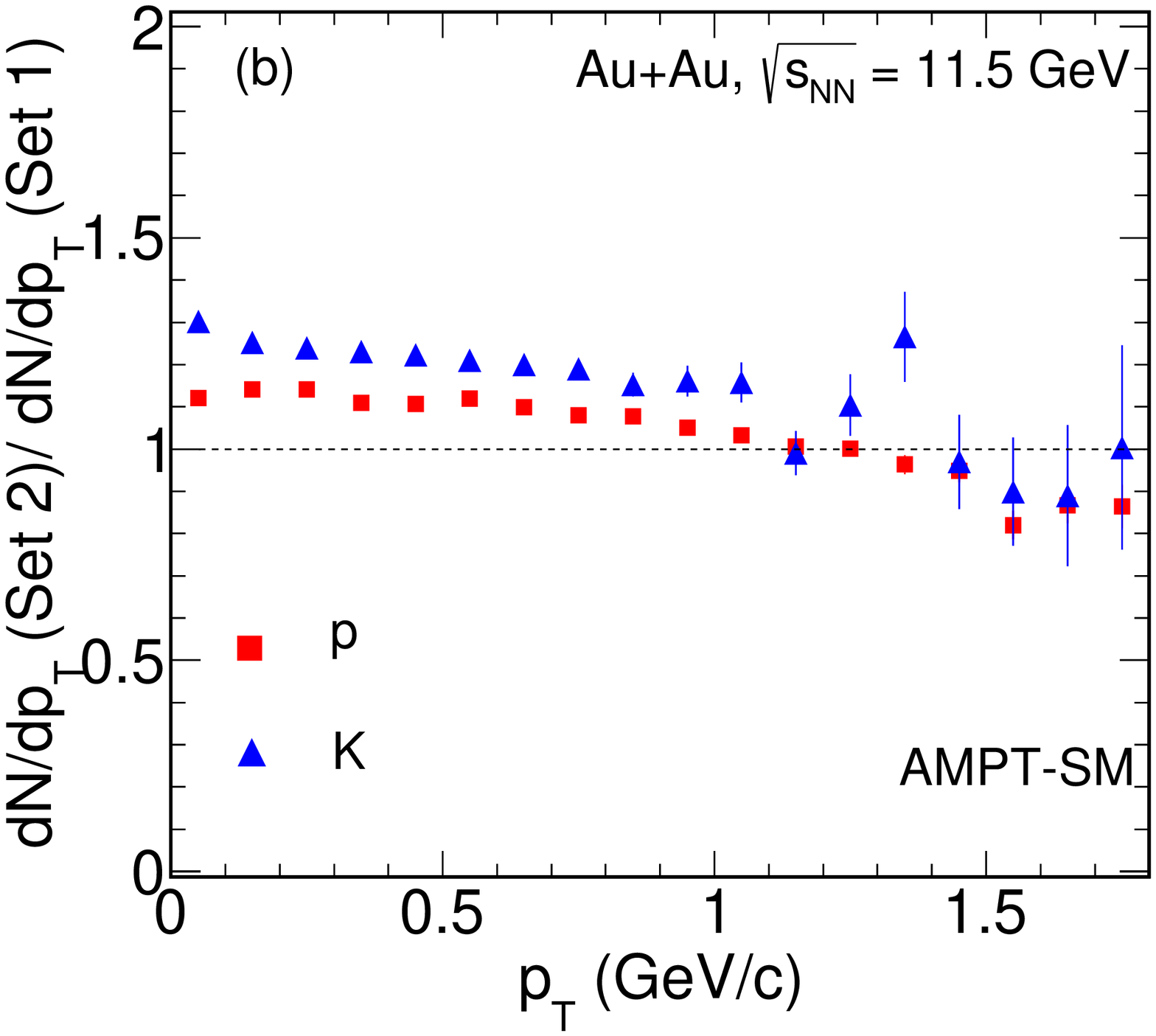}
\caption{(Color online)  (a) Ratios of pions and $\phi$ mesons $p_{T}$ spectra at mid-rapidity between two sets of input parameters of AMPT in Au+Au collisions at $\sqrt{s_{\mathrm {NN}}}$ = 11.5 GeV.  (b)  Ratios of kaons and protons $p_{T}$ spectra at mid-rapidity between two sets of input parameters of AMPT in Au+Au collisions at $\sqrt{s_{\mathrm {NN}}}$ = 11.5 GeV. Here set 1 corresponds to   $a$=0.55 and $b$ =0.15 and set 2 corresponds to $a$=2.2 and $b$ =0.15.}
\label{fig1}
\end{center}
\eef
The values of Lund string fragmentation parameters ($a$ and $b$) were determined by fitting the charged particle yield and $p_{T}$ spectra measured in the experiment for different collisions system~\cite{ampt,lund_para1,lund_para2}. In this work, we included the production of strange quarks carrying hadrons, $\phi$ meson, as well as non-strange $\pi$ meson to determine the Lund parameters, $a$ and $b$, for Au+Au collisions at  $\sqrt{s_{\mathrm {NN}}}$ = 11.5, 39,  and 200 GeV. We have used two sets for parameters; set 1:  $a$=0.55 and $b$ =0.15 (proposed in ref.~\cite{lund_para1}) and  set 2 : $a$=2.2 and $b$ =0.15.
The parton-parton cross-section is taken to be 3 mb (using strong coupling constant $\alpha_{S}$ =0.33) for both set 1 and set 2.    The termination time of the hadronic cascade is taken to be 30 $fm/c$. \\
In the public version of the AMPT model, the deformation parameter of Uranium nucleus is not implemented. In this work, we have implemented deformed shape of the Uranium nucleus in the AMPT model to study the U+U collisions.
The nucleon density distribution is parametrized as a deformed Woods-Saxon profile. 
\be
\rho = \frac{\rho_{0}}{1+exp([r-R]/a)},
\ee
\be
R=R_{0}[1+\beta_{2}Y^{0}_{2}(\theta) + \beta_{4}Y^{0}_{4}(\theta)],
\ee

Where $\rho_{0}$ is the normal nuclear density and $Y^{l}_{m} (\theta )$ denote spherical harmonics and $\theta$ is the polar angle with respect to the symmetry axis of the nucleus. 
We have used the radius of the nucleus $R_{0}$ = 6.81 $fm$ and the surface diffuseness parameter $a$= 0.55 $fm$ for U nucleus.   The deformation parameters in the Woods–Saxon profile are $\beta_{2}$= 0.28 and $\beta_{4}$=0.093~\cite{uu1,uu2}.

\section{Results}
\subsection{Au+Au collisions}
 We have calculated the transverse momentum spectra of pion and $\phi$ meson using the AMPT model with two sets of parameters, set 1 and set 2.
Figure~\ref{fig1} (a) shows the ratios of pions and $\phi$ mesons spectra between two sets of input parameters in AMPT. The solid red circle and solid blue star marker  correspond to $\pi^{-}$ and $\phi$ mesons, respectively.
This calculation is done for 10-40\% Au+Au collisions at $\sqrt{s_{\mathrm {NN}}}$ = 11.5 GeV. From Fig.~\ref{fig1}, we can see that change in $\pi$ spectra is within 15\%, however, a large difference ($\sim$50\% at low $p_{T}$)  is observed for $\phi$ meson spectra. A similar comparison has been shown for kaons (consist of one strange quark) and protons (similar mass as $\phi$) in Fig~\ref{fig1} (b), where a change in spectra is found to be $\sim$ 20-25\%  and 10-15\%, respectively. This study suggests that yield of strange quarks carrying hadrons are more sensitive observable to constraint the Lund fragmentation parameter. These observations are energy independent. 
In this paper, we have used $p_{T}$ spectra of pions and $\phi$ meson simultaneously to extract Lund fit parameters for Au+Au collisions at $\sqrt{s_{\mathrm {NN}}}$ = 11.5, 39 and 200 GeV.

\bef
\begin{center}
\includegraphics[scale=0.36]{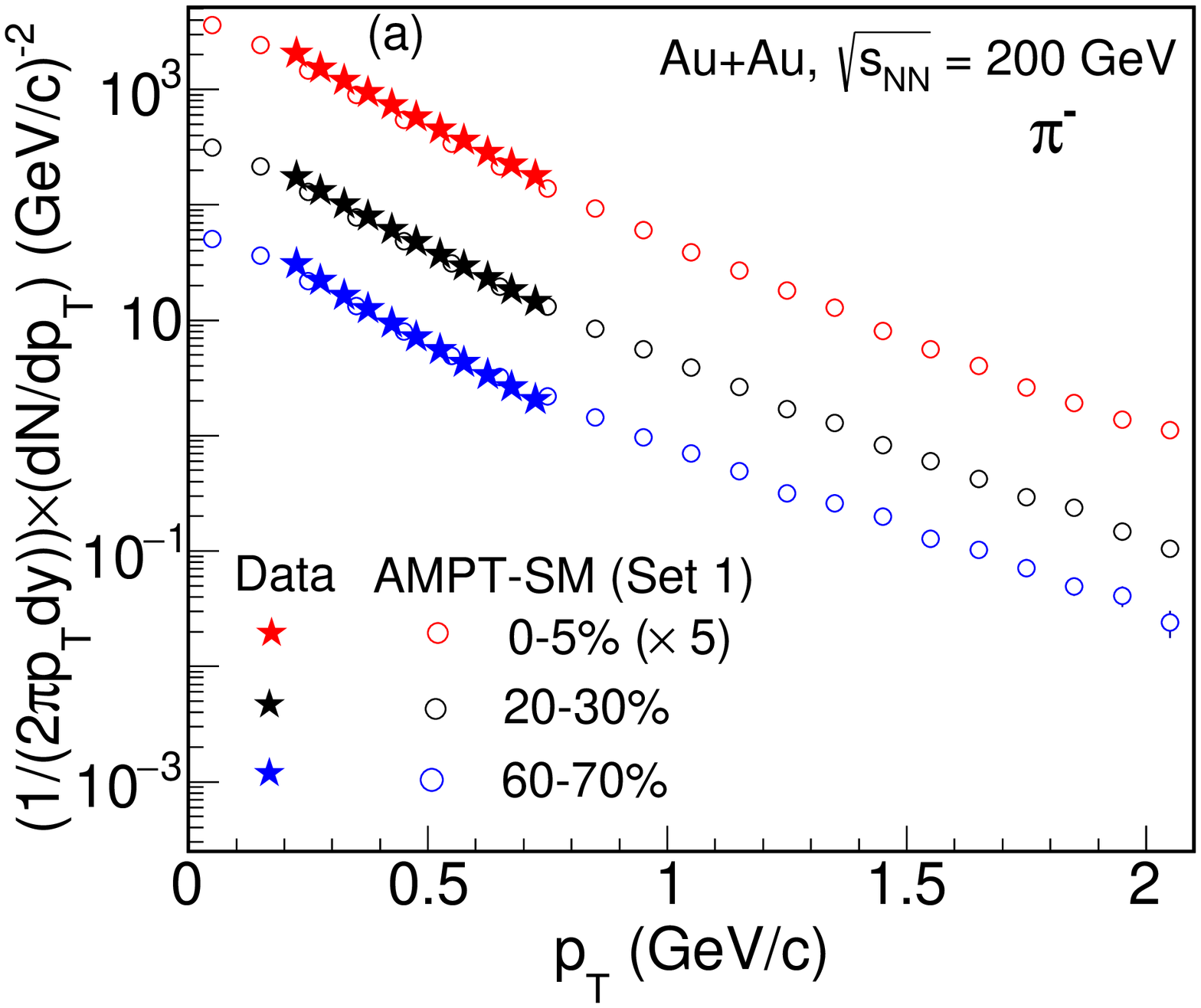}
\includegraphics[scale=0.36]{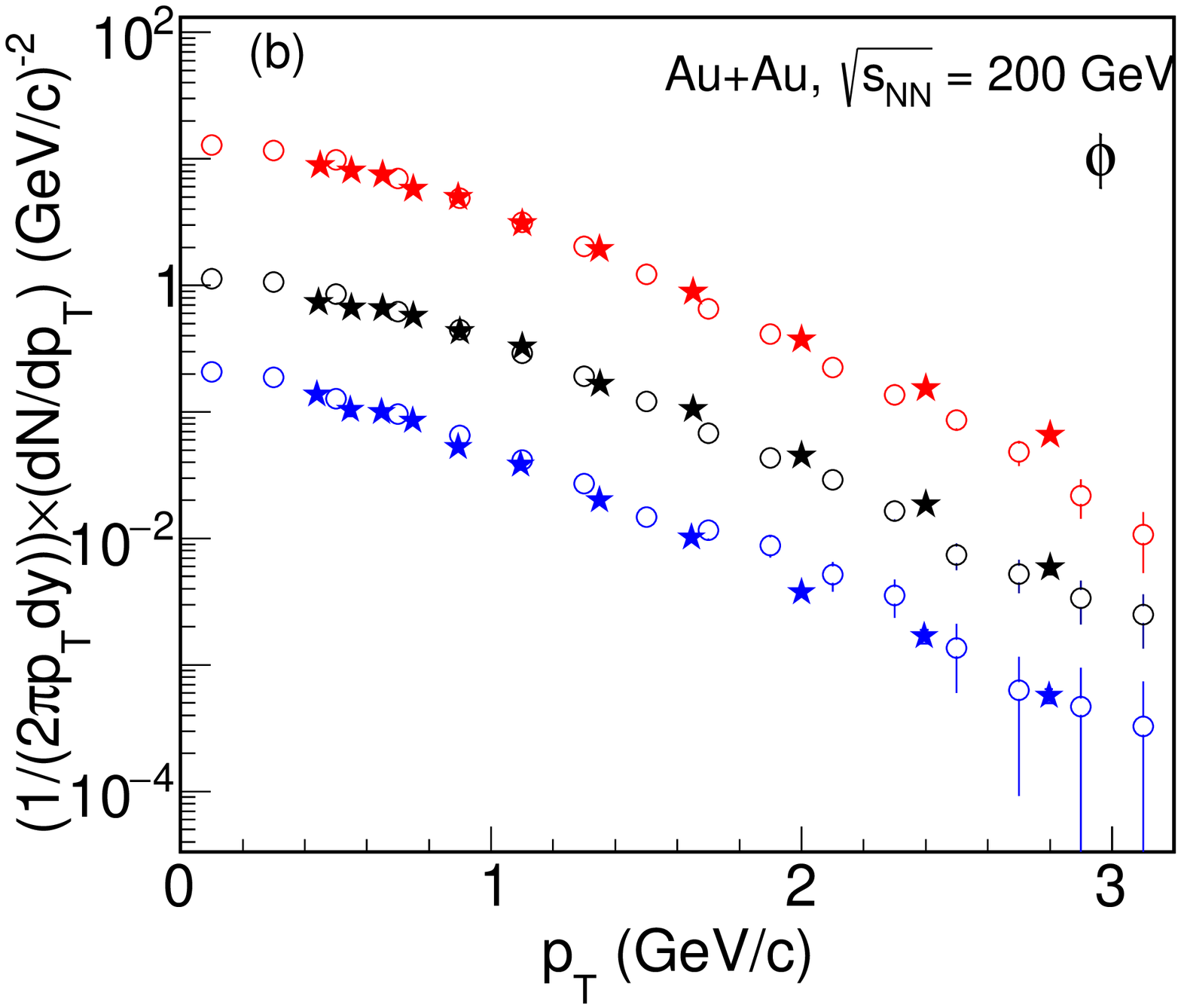}
\caption{(Color online) Invariant yield of (a) $\pi^{-}$  and (b) $\phi$  at mid-rapidity as a function of transverse momentum in Au+Au collisions at  $\sqrt{s_{\mathrm {NN}}}$ = 200 GeV  for 0-5\%, 20-30\% and 60-70\% centralities. Solid star marker represents experimentally measured values~\cite{pi_200gev,phi_200gev} and open circle represents AMPT model calculation using  $a$=0.55 and $b$ =0.15 (set 1).}
\label{fig2}
\end{center}
\eef
 
 Figure~\ref{fig2}(a) and (b) compare the experimentally measured $p_{T}$ spectra of non-strange pion and $\phi$ meson with AMPT model calculation  in Au+Au collisions at  $\sqrt{s_{\mathrm {NN}}}$ = 200 GeV~\cite{pi_200gev,phi_200gev}.
The measurement has been done at mid-rapidity, $|y|$ $<$ 0.1 (for $\pi$) and $|y|$ $<$ 0.5 (for $\phi$).  Comparison is shown for three different centralities, 0-5\% (most central), 20-30\% (mid-central) and 60-70\% (peripheral). 
Lund parameters ($a$ = 0.55 and $b$ = 0.15), used to fit charged particle spectra in Au+Au collisions at  $\sqrt{s_{\mathrm {NN}}}$ = 200 GeV~\cite{lund_para1},  are also used here for both pions and $\phi$ meson. We can see AMPT model calculations, with Lund parameters $a$=0.55 and $b$ =0.15, explain both non-strange $\pi^{-}$ and strange $\phi$ meson spectra for all $p_{T}$ and all centralities.

After observing good agreement between data and model  using parameter  $a$=0.55 and $b$ =0.15 (set 1), we wanted to check what happens if we lower the center-of-mass energy. In Fig.~\ref{fig3}, we have shown comparison of $\pi^{-}$ and $\phi$ meson spectra between data and AMPT model using   parameter $a$=0.55 and $b$ =0.15 (set 1) at $\sqrt{s_{\mathrm {NN}}}$ = 39 GeV.  We can see, in most central  and mid-central collisions AMPT model explains the data very well for both $\pi^{-}$ and $\phi$ mesons. However, model over predicts the data at high $p_{T}$ for most peripheral collisions.

Figure~\ref{fig4} shows comparison of $\pi^{-}$ and $\phi$ meson spectra between data and AMPT model  for Au+Au collisions at $\sqrt{s_{\mathrm {NN}}}$ = 11.5 GeV. Here we can see, Lund parameter $a$=0.55 and $b$ =0.15 (set 1) explains $\pi^{-}$ spectra at most and mid-central collisions. However, AMPT model calculation using Lund parameter $a$=0.55 and $b$ =0.15 (set 1) under-predicts measured $\phi$ meson spectra at most and mid-central collisions. We have also shown AMPT model results using Lund parameter $a$=2.2 and $b$ =0.15 (set 2) in Fig.~\ref{fig4}. We do not see any significant change in $\pi^{-}$ spectra; however, the comparison between data and AMPT model with set 2 parameter looks better than that of set 1 for $\phi$ meson. This clearly shows $\phi$ meson is a better probe to have control over the Lund parameter. It is important to note that, using the same set of parameters (set 1), one can explain data at both 200 and 39 GeV, however. we need a different set of parameters in order to describe data at 11.5 GeV. This suggests that the underlying mechanism for particle production at 11.5 GeV  could be different than that of at 39 and 200 GeV. This finding is consistent with the finding by the STAR experiment as presented in ref.~\cite{phi_omega_star}.

\bef
\begin{center}
\includegraphics[scale=0.36]{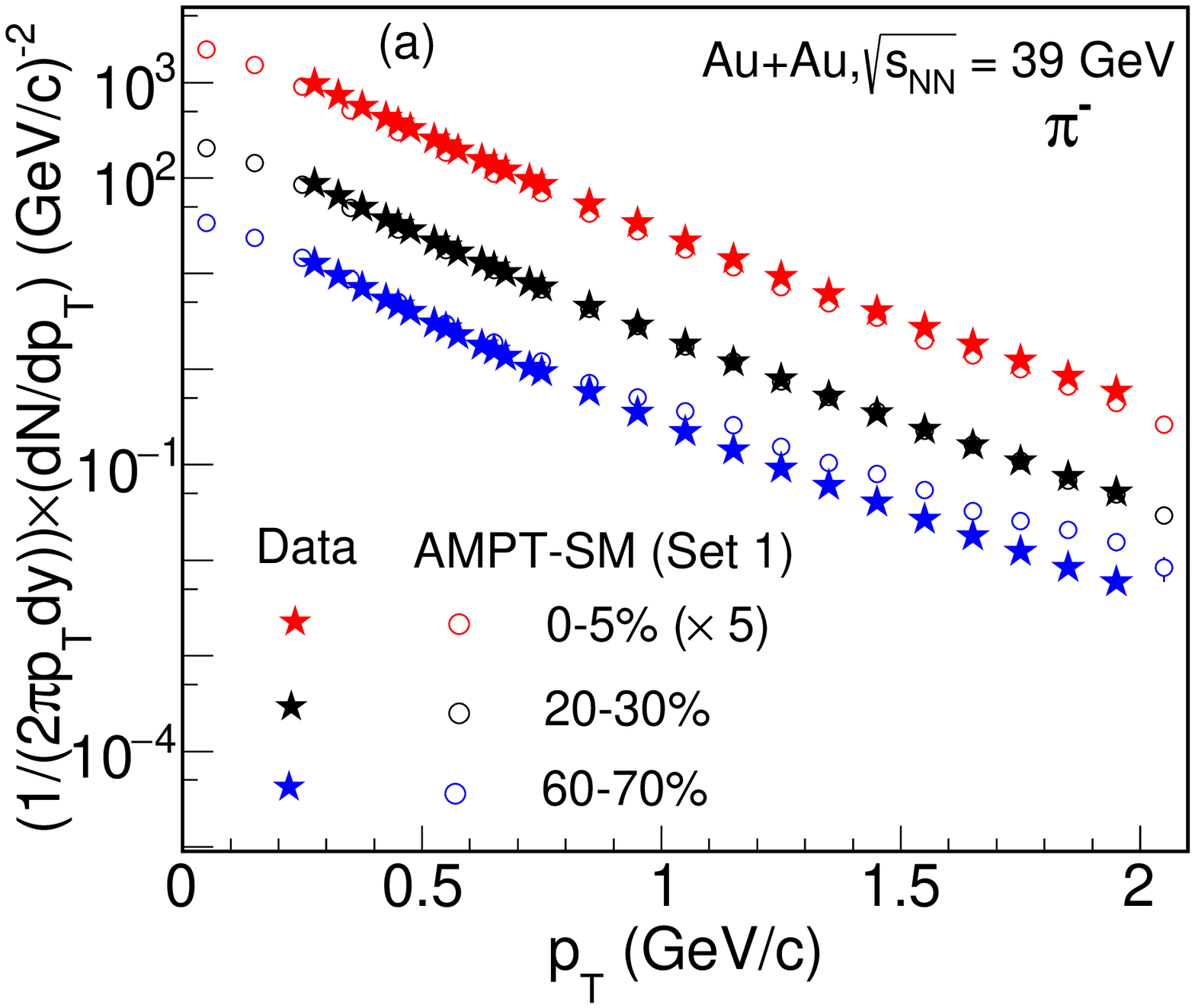}
\includegraphics[scale=0.36]{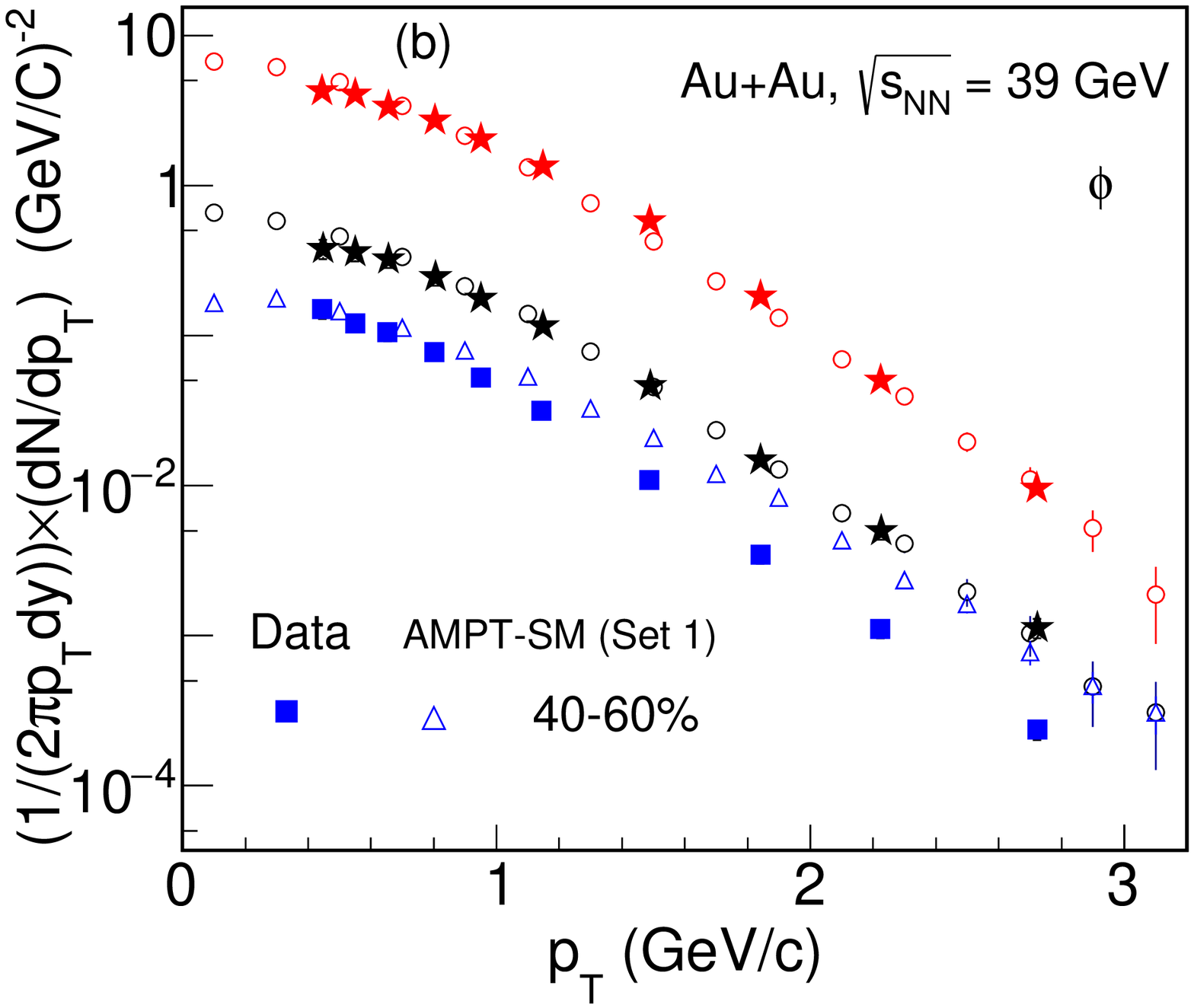}
\caption{(Color online) Invariant yield of (a) $\pi^{-}$.  and (b) $\phi$  at mid-rapidity as a function of transverse momentum in Au+Au collisions at  $\sqrt{s_{\mathrm {NN}}}$ = 39 GeV  for 0-5\%, 20-30\% and 60-70\% (40-60\% for $\phi$) centralities. Solid star marker represents experimentally measured values~\cite{pikp_star,phi_omega_star} and open circle represents AMPT model calculation using  $a$=0.55 and $b$ =0.15 (set 1)}
\label{fig3}
\end{center}
\eef

\bef
\begin{center}
\includegraphics[scale=0.36]{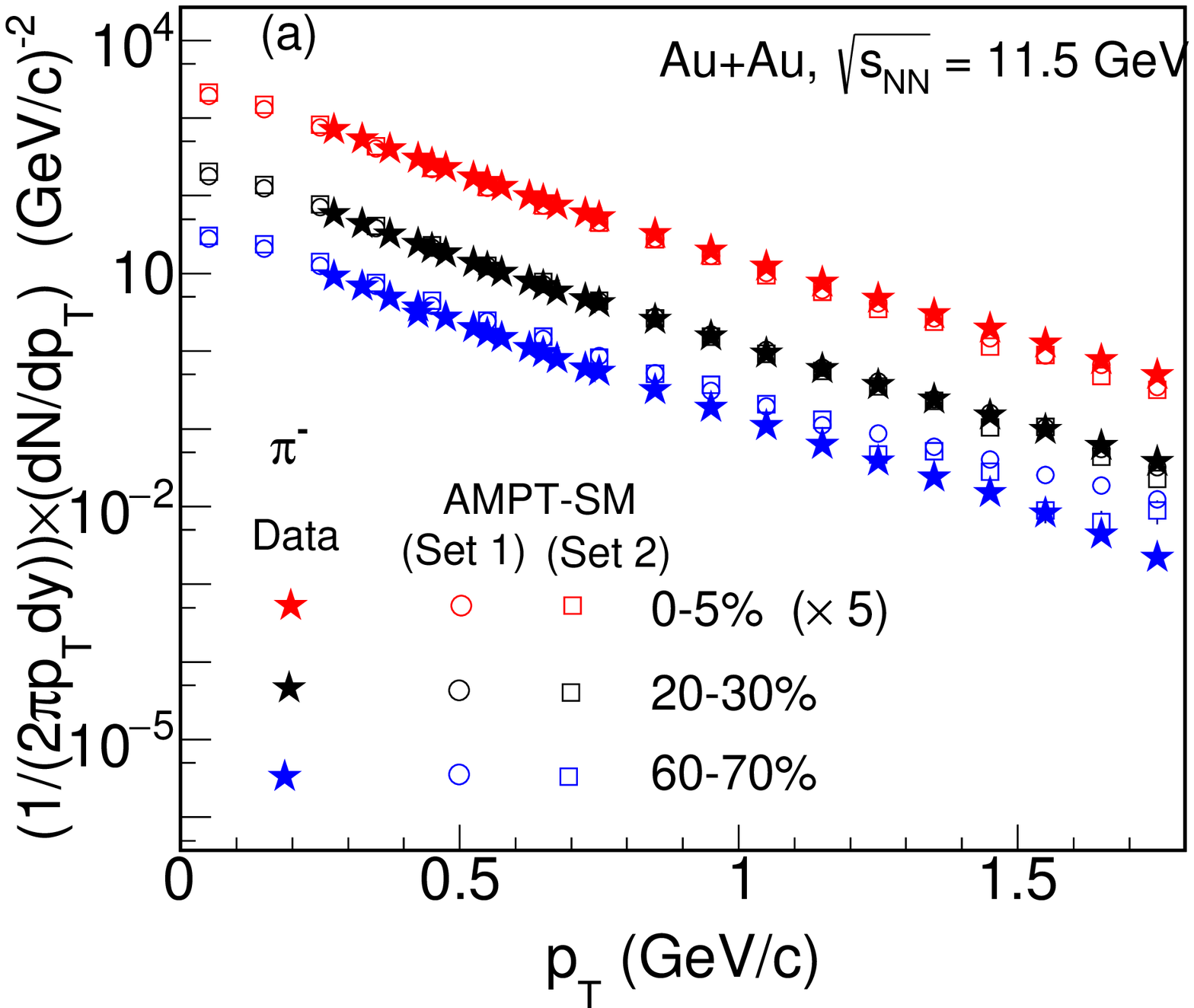}
\includegraphics[scale=0.36]{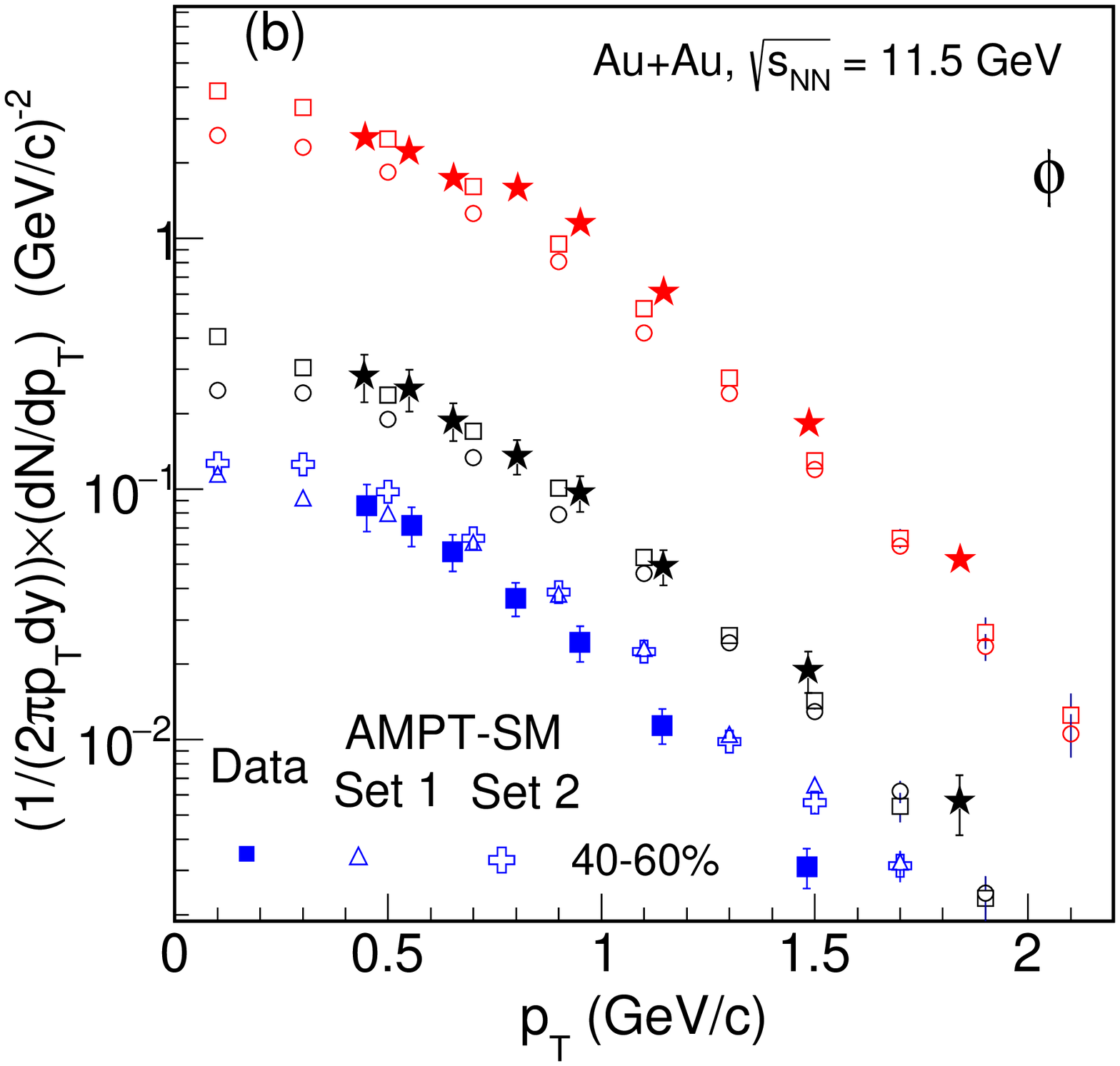}
\caption{(Color online) Invariant yield of (a) $\pi^{-}$.  and (b) $\phi$ at mid-rapidity as a function of transverse momentum in Au+Au collisions at  $\sqrt{s_{\mathrm {NN}}}$ = 11.5 GeV  for 0-5\%, 20-30\% and 60-70\% (40-60\% for $\phi$) centralities. Solid star marker represents experimentally measured values~\cite{pikp_star,phi_omega_star} and open circle represents AMPT model calculation using  parameters set 1 ($a$=0.55 and $b$ =0.15) and set 2 ($a$=2.2 and $b$ =0.15)}
\label{fig4}
\end{center}
\eef

 \subsection{U+U collisions}
After obtaining suitable sets of parameters of the AMPT model in Au+Au collisions at top RHIC and BES-energies, we have calculated the invariant yield of $\pi^{-}$  and  $\phi$ mesons spectra in U+U collisions  at  $\sqrt{s_{\mathrm {NN}}}$ = 196 GeV  for various centralities. The U+U collisions are expected to produce more extreme conditions of QCD matter at higher density and/or greater volume than  the spherical gold nuclei at the same incident energy.  Since Uranium has quadrupole deformed shape, the collisions between Uranium nuclei will produce different initial geometry compared to Au+Au collisions. 
Figure~\ref{fig5} shows the Invariant yield of (a) $\pi^{-}$  and (b) $\phi$ at mid-rapidity as a function of transverse momentum in U+U collisions at  $\sqrt{s_{\mathrm {NN}}}$ = 196 GeV  for 0-5\%, 20-30\% and 60-70\% centralities.  We also compared the results with that of  Au+Au collisions at  $\sqrt{s_{\mathrm {NN}}}$ = 200 GeV. In this calculation, we have used  Lund parameters $a$=0.55 and $b$ =0.15 (set 1)  which explain data for Au+Au collisions at  $\sqrt{s_{\mathrm {NN}}}$ = 200 GeV. It is found that the invariant yield of $\pi^{-}$  and $\phi$ is higher (20$\%$) in U+U collisions compared to that in Au+Au collisions. This could be due to the higher initial energy density in U+U collisions compared to Au+Au collisions. These AMPT results can be compared to the experimental results to be available soon from STAR collaboration. 

\bef
\begin{center}
\includegraphics[scale=0.36]{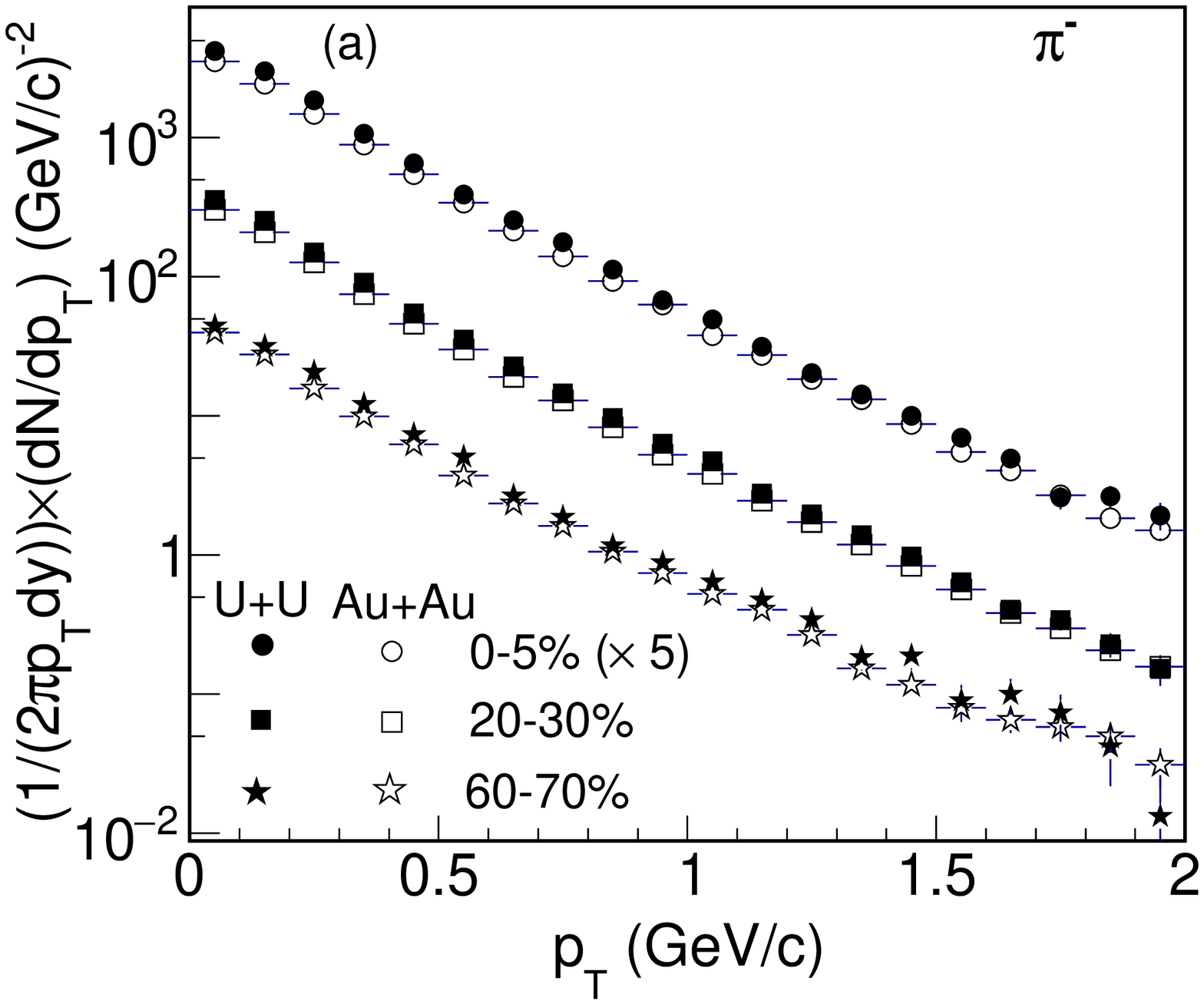}
\includegraphics[scale=0.36]{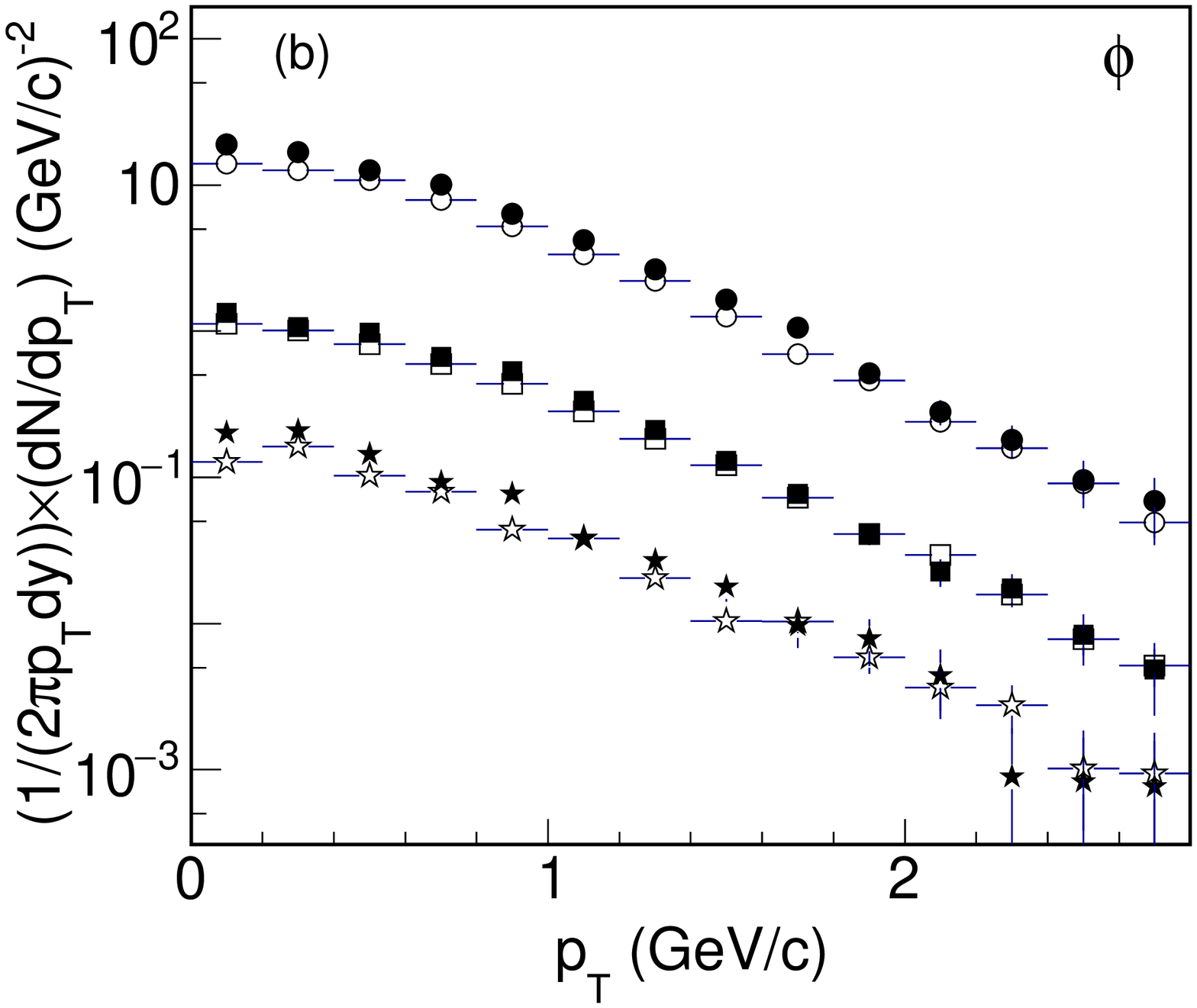}
\caption{(Color online) Invariant yield of (a) $\pi^{-}$  and (b) $\phi$ at mid-rapidity as a function of transverse momentum in U+U collisions at  $\sqrt{s_{\mathrm {NN}}}$ = 196 GeV  for 0-5\%, 20-30\% and 60-70\% centralities. The AMPT model calculation using  parameters set 1 ($a$=0.55 and $b$ =0.15) which explains data for Au+Au collisions at  $\sqrt{s_{\mathrm {NN}}}$ = 200 GeV. The AMPT model calculation for Au+Au collisions at  $\sqrt{s_{\mathrm {NN}}}$ = 200 GeV using same set of parameters are also shown for comparison.}
\label{fig5}
\end{center}
\eef

\section{Summary}

In summary, a compilation of the available data at RHIC energies for the invariant yield  of $\pi^{-}$ and  $\phi$ mesons has been presented. We have shown that strange quarks carrying hadrons are the most sensitive probe to the initial parton dynamics and one should use measured $\phi$ meson yield along with non-strange particle to determine the model parameters in the AMPT model.  It has been found that using the same set of parameters; one can explain both $\pi^{-}$ and $\phi$ meson spectra at 200 and 39 GeV. However, one needs a different set of input parameters to explain measured $\phi$ meson spectra at 11.5 GeV.  These results suggest that there is likely a change in the underlying strange quark dynamics in the produced matter during Au+Au collisions. In addition, we have studied the production of $\pi$ and $\phi$ meson in U+U collisions  at  $\sqrt{s_{\mathrm {NN}}}$ = 196 GeV after implementing deformation for the uranium nucleus in the AMPT model. These results can be compared to experimental  measurements at RHIC to understand the particle production mechanism in U+U collisions.



\end{document}